\documentclass{article}

\usepackage{arxiv}

\usepackage[utf8]{inputenc} % allow utf-8 input
\usepackage[T1]{fontenc}    % use 8-bit T1 fonts
\usepackage{hyperref}       % hyperlinks
\usepackage{url}            % simple URL typesetting
\usepackage{booktabs}       % professional-quality tables
\usepackage{amsfonts}       % blackboard math symbols
\usepackage{nicefrac}       % compact symbols for 1/2, etc.
\usepackage{microtype}      % microtypography
\usepackage{lipsum}
\usepackage{amsmath}
\usepackage{amssymb}
\usepackage{graphicx}
\usepackage{psfrag}

\newcommand{\be}{\begin{eqnarray}}
\newcommand{\ee}{\end{eqnarray}}

\newcommand{\bfp}{{\bf p}_{\perp}}

\title{Parton Distribution Functions for diquarks based on an AdS/QCD quark-diquark nucleon model}

\author{
  Benjamin Rodriguez-Aguilar\thanks{ \texttt{rodrigesagilar.l@edu.spbstu.ru}} ~and~ Ya. A. Berdnikov\thanks{ \texttt{berdnikov@spbstu.ru}} \\% \thanks{Use footnote for providing further     information about author (webpage, alternative    address)---\emph{not} for acknowledging funding agencies.} \\
  Peter the Great St.  Petersburg Polytechnic University,\\
  Polytechnicheskaya 29, 195251, Russian Federation\\
}

\begin{document}
\maketitle

\begin{abstract}
%\lipsum[1]
We present a phenomenological unpolarized Parton Distribution Functions for diquarks based on a soft-wall light front AdS/QCD quark-diquark nucleon model. From a probed model consistent with the Drewll-Yan-West relation and quark counting rule, we have performed a fit of some free parameters using known phenomenological quark PDF data. The model considers the entire set of possible diquarks within the nucleon valence, in the present work we focus on the spin-0 $ud_0$, spin-1 $ud_1$ and spin-1 $uu_1$ diquarks into the valence of protons.   The diquark PDFs obtained are able to used in proton-proton collision simulations.% in parallel processes
%able to be used in particle collision simulations. With the construction of early 
\end{abstract}

% keywords can be removed

\section{Introduction}
%\lipsum[2]
%\lipsum[3]
Since the end of the last decade of the XX century, the AdS/CFT correspondence \cite{Maldacena:1997re} between string theory in Anti-de-Sitter (AdS) space-time and conformal field theories (CFT) in physical space-time has been a very active and interesting field of study. Among other things, the wealth of this correspondence stands in the possibility to perform calculations between opposite coupling regimes, strongly coupled theories can be mapped into weakly coupled ones and vice-versa. CFTs are defined as scale invariant theories, so is not possible a directly application of the AdS/CFT correspondence to QCD itself. In fact, is because the coupling constants change with the renormalization scale $\mu$ in QCD that we get the condition under which perturbation theory is valid \cite{Peskin:1995ev}.

Nevertheless,  in the strong coupling regime of QCD, the couplings appear to be approximately constant constant. This is the basis for a Light-front holography, an approximation of the AdS/CFT to QCD (light-front AdS/QCD) \cite{deTeramond:2008ht} that has shown the ability to find analytic solutions int the non-perturvative regime of QCD, like improving predictions of hadron masses and structure properties, see eg. \cite{Brodsky:2007hb}. 

In this work, we are particularly interested in the fact that light-front AdS/QCD predicts a general form of two particle bound state wave function inside nucleons which can not be derived simply from valence quarks \cite{Brodsky:2007hb,Maji:2016yqo}. This has led to considerable progress in nucleon analytical results considering valence diquarks in their structure, just as  light-front wave functions QCD matched with soft-wall AdS/QCD predictions \cite{Gutsche:2011vb, Gutsche:2013zia, Bacchetta:2008af}. Another recent result contemplates the scale evolution of  the  Parton Distribution Functions (PDFs)  for a quark-diquark nucleon model using scale dependent parameters following the DGLAP evolution \cite{Maji:2016yqo}, that are consistent with the quark counting rule and Drell-Yan-West relation \cite{Drell:1969km, West:1970av}. Based on these last two results, we have fitted the PDF parameters of the quark-diquark  nucleon model to the available data from NNPDF2.3 QCD+QED NNLO  \cite{Ball:2013hta} for $u$ and $d$ quarks, in order to get the unpolarized PDFs for the spin-0 $ud_0$, spin-1 $ud_1$ and spin-1 $uu_1$ diquarks. 
With such parameters available, the diquark PDFs can be used to simulations of proton (and neutron) collisions with participating  diquarks.

To consider proton collisions based on a nucleon model with diquark structures inside, it is useful to inspect the properties of the parton model. The cross section for proton-proton collisions can be expressed by the so called \textit{improved parton model formula} \cite{Nason:2001mq}
{%\small
\begin{equation}\label{pdfformula}
    \sigma_{(P_1,P_2)} = \sum_{i,j} \int dx_1 dx_2 f^{1}_i (x_1,\mu) f^{2}_j (x_2,\mu)~\hat{\sigma}_{ij} (x_1 P_1, x_2 P_2, \alpha_s (\mu), \mu),
\end{equation}
}
where the scripts $1,2$ are labels to the incoming protons beams carried momentum $P$. In this scenario, the incoming proton beam is equivalent to a beam made of constituent \textit{partons}. Typically, these partons are taken as the masless-pointlike elementary particles, quarks and gluons \cite{Nason:2001mq}, with a longitudinal momentum distribution characterized by the Parton Distribution Functions $f_i(x,\mu)$. This means, given some proton with momentum $P$, the probability to find in such proton the parton $i$ with momentum between $xP$ and $(x + dx)P$ is precisely $dx f_i(x,\mu)$, being dependent as well of the renormalization scale $\mu$. While $\hat{\sigma}$ represents the parton cross sections, which can be computed with perturbative QCD (pQCD) for sufficiently small running coupling $\alpha_s (\mu)$.\\
However, due to the fact that (coloured) partons cannot be observed as free particles, the PDFs cannot be calculated using pQCD. Nowadays, the simplest way to obtain PDFs is fitting observables to experimental data, among other phenomenological tools, see e.g. \cite{Pumplin:2002vw,Ball:2014uwa}.

Nevertheless, in order to work with a parton model using constituent diquarks, we must to expand this picture beyond quarks and gluons. Recent results from soft-wall AdS/QCD \cite{Brodsky:2007hb, Gutsche:2013zia} have shown a phenomenological approach to reproduce unpolarized PDFs of quark-diquark nucleons \cite{Maji:2016yqo}. In next section we show how this phenomenological approach has been constructed to finally obtain our parameters that allow us to exhibit our diquark PDFs.
%On the other hand, the diquark cross sections can be approached using the color symmetry in hadrons and introducing a form factor. Such procedure is described in the next sections.

\section{The soft-wall light front AdS/QCD quark-diquark nucleon model}
\label{sec:nucleonmodel}
%\lipsum[4] See Section \ref{sec:headings}.
In this section we intend to outline how to obtain the PDF of a quark-diquark nucleon model using soft-wall light front holographic QCD. For a more detailed analysis see \cite{Maji:2016yqo} (and its references), from where this section is heavily based.\\

To construct such a PDF model, it is assumed that a virtual incoming photon interacts with a massless-valence quark. The other two valence quarks are then forming a spectator diquark. In this way, it is ensured that this model is in accordance with the traditional quark-interacting frameworks, from where it is possible to build reliable properties for the nucleon model, so for diquarks.
The diquarks can have then either spin-0 (scalar diquark) or spin-1 (vector diquark).\\
The nucleon state is represented by a spin-flavour SU(4) symmetry. 
This implies that the possible states are the isoscalar-scalar diquark singlet state, the isoscalar-vector diquark state and the isovector-vector diquark state. Shortly, the diquark can be either scalar or axial-vector.  For the proton state we can write it as
%The spin-0 diquarks are always in a flavour singlet state (isoscalar-scalar diquark), while the spin-1 diquarks are either in  flavour (isoscalar-vector diquark) or in 
\begin{equation}\label{PS_state}
   |P; \pm\rangle = C_S|u~ S^0\rangle^\pm + C_V|u~ A^0\rangle^\pm + C_{VV}|d~ A^1\rangle^\pm,
\end{equation}
where, following the original notation in \cite{Maji:2016yqo}, $S$ and $A$ represent the scalar and vector diquark having isospin at their superscript. The subscript in the coefficients denotes the isoscalar-scalar ($S$), the isoscalar-vector state ($V$) and the isovector-vector state ($VV$). For the neutron state is given by the isospin symmetry $u\leftrightarrow d$. Without losing the generality of the model, we will take the case for the proton, which is what we care about in this work.\\

Using the light-cone convention $x^\pm=x^0 \pm x^3$ \cite{Lepage:1980fj}, it is convenient to choose a frame where the proton transverse momentum vanishes, denoted as $P \equiv \big(P^+,\frac{M^2}{P^+},\textbf{0}_\perp\big)$, where $M$ is the proton mass. So the momentum of the struck quark can be taken as $p\equiv (xP^+, \frac{p^2+|\bfp|^2}{xP^+},\bfp)$  and  the diquark,   $P_X\equiv ((1-x)P^+,P^-_X,-\bfp)$. We can interpret from this notation that $x=p^+/P^+$ is the longitudinal momentum fraction carried by the struck quark. \\

Now, we can express the two particle Fock-state expansion. For $J^z =\pm1/2$ with spin-0 diquark is given by
\begin{equation}
    |u~ S\rangle^\pm  = \int \frac{dx~ d^2\bfp}{2(2\pi)^3\sqrt{x(1-x)}} \bigg[ \psi^{\pm(u)}_{+}(x,\bfp)|+\frac{1}{2}~s; xP^+,\bfp\rangle 
 + \psi^{\pm(u)}_{-}(x,\bfp)|-\frac{1}{2}~s; xP^+,\bfp\rangle\bigg],\label{fock_PS}
\end{equation}
where $|\lambda_q~\lambda_S; xP^+,\bfp\rangle$ is the two particle state having struck quark of helicity $\lambda_q$ and a scalar diquark having helicity $\lambda_S=s$ (spin-0 singlet diquark helicity is denoted by $s$ to distinguish from triplet diquark). While, the spin-1 diquark state is given by \cite{Ellis:2008in}
\be
|\nu~ A \rangle^\pm & =& \int \frac{dx~ d^2\bfp}{2(2\pi)^3\sqrt{x(1-x)}} \bigg[ \psi^{\pm(\nu)}_{++}(x,\bfp)|+\frac{1}{2}~+1; xP^+,\bfp\rangle \nonumber\\
 &+& \psi^{\pm(\nu)}_{-+}(x,\bfp)|-\frac{1}{2}~+1; xP^+,\bfp\rangle +\psi^{\pm(\nu)}_{+0}(x,\bfp)|+\frac{1}{2}~0; xP^+,\bfp\rangle \nonumber \\
 &+& \psi^{\pm(\nu)}_{-0}(x,\bfp)|-\frac{1}{2}~0; xP^+,\bfp\rangle + \psi^{\pm(\nu)}_{+-}(x,\bfp)|+\frac{1}{2}~-1; xP^+,\bfp\rangle \nonumber\\
 &+& \psi^{\pm(\nu)}_{--}(x,\bfp)|-\frac{1}{2}~-1; xP^+,\bfp\rangle  \bigg],\label{fock_PA}
\ee
where $|\lambda_q~\lambda_D; xP^+,\bfp\rangle$ represents a two-particle state with a quark of helicity $\lambda_q=\pm\frac{1}{2}$ and a vector diquark of helicity $\lambda_D=\pm 1,0$ (triplet). Here $\nu = u,d$ is a flavour index.\\

The light-front (LF) wave functions with spin-0 diquark state, $\psi^{\pm~(u)}_{\pm}$, at the initial scale $\mu_0$ for $J=\pm1/2$ are given by \cite{Lepage:1980fj}
\begin{equation}\label{LFWF_S1}
  J = +1/2 :
    \begin{cases}
      \psi^{+(u)}_+(x,\bfp) =& N_S~ \varphi^{(u)}_{1}(x,\bfp),\\
\psi^{+(u)}_-(x,\bfp) =& N_S\bigg(- \frac{p^1+ip^2}{xM} \bigg)\varphi^{(u)}_{2}(x,\bfp),
    \end{cases}       
\end{equation}
\begin{equation}\label{LFWF_S2}
  J = -1/2 :
    \begin{cases}
      \psi^{-(u)}_+(x,\bfp)=& N_S \bigg(\frac{p^1-ip^2}{xM}\bigg) \varphi^{(u)}_{2}(x,\bfp), \\
\psi^{-(u)}_-(x,\bfp)=&  N_S~ \varphi^{(u)}_{1}(x,\bfp).
    \end{cases}       
\end{equation}

In a very similar way, for vector diquarks with  $J=\pm1/2$ the LF wave functions $\psi^{\pm~(\nu)}_{\pm~\pm}$ at the initial scale $\mu_0$ can be written as
\begin{equation}\label{LFWF_Vp}
  J = +1/2 :
    \begin{cases}
      \psi^{+(\nu)}_{+~+}(x,\bfp)= N^{(\nu)}_1 \sqrt{\frac{2}{3}} \bigg(\frac{p^1-ip^2}{xM}\bigg) \varphi^{(\nu)}_{2}(x,\bfp),\hspace{0.2cm}&
\psi^{+(\nu)}_{-~+}(x,\bfp)= N^{(\nu)}_1 \sqrt{\frac{2}{3}} \varphi^{(\nu)}_{1}(x,\bfp), \\
      \psi^{+(\nu)}_{+~0}(x,\bfp)= - N^{(\nu)}_0 \sqrt{\frac{1}{3}} \varphi^{(\nu)}_{1}(x,\bfp),\hspace{0.2cm}&
\psi^{+(\nu)}_{-~0}(x,\bfp)= N^{(\nu)}_0 \sqrt{\frac{1}{3}} \bigg(\frac{p^1+ip^2}{xM} \bigg)\varphi^{(\nu)}_{2}(x,\bfp), \\
\psi^{+(\nu)}_{+~-}(x,\bfp) = 0, \hspace{0.2cm}&
\psi^{+(\nu)}_{-~-}(x,\bfp) =  0, 
    \end{cases}       
\end{equation}
\begin{equation}\label{LFWF_Vm}
  J = -1/2 :
    \begin{cases}
      \psi^{-(\nu)}_{+~+}(x,\bfp)= 0,\hspace{0.2cm}&
\psi^{-(\nu)}_{-~+}(x,\bfp)= 0, \\
\psi^{-(\nu)}_{+~0}(x,\bfp)= N^{(\nu)}_0 \sqrt{\frac{1}{3}} \bigg( \frac{p^1-ip^2}{xM} \bigg) \varphi^{(\nu)}_{2}(x,\bfp),\hspace{0.2cm}&
\psi^{-(\nu)}_{-~0}(x,\bfp)= N^{(\nu)}_0\sqrt{\frac{1}{3}} \varphi^{(\nu)}_{1}(x,\bfp), \\
\psi^{-(\nu)}_{+~-}(x,\bfp)= - N^{(\nu)}_1 \sqrt{\frac{2}{3}} \varphi^{(\nu)}_{1}(x,\bfp),\hspace{0.2cm}&
\psi^{-(\nu)}_{-~-}(x,\bfp)= N^{(\nu)}_1 \sqrt{\frac{2}{3}} \bigg(\frac{p^1+ip^2}{xM}\bigg) \varphi^{(\nu)}_{2}(x,\bfp).
    \end{cases}       
\end{equation}
The LF wave functions $\varphi_i^\nu ~(i=1,2)$ are the twist-3 LF wave functions. These funtions can be derived in light-front QCD and in soft-wall AdS/QCD \cite{Brodsky:2007hb,Brodsky:2011xx,Abidin:2009hr,Vega:2009zb,Gutsche:2011vb}. In \cite{Gutsche:2013zia} has been proposed a generalized form to $\varphi_i^\nu$ by matching the electromagnetic form factors of the nucleon in soft-wall AdS/QCD and light-front QCD, getting that
\begin{equation}
    \varphi_i^{(\nu)}(x,\bfp)=\frac{4\pi}{\kappa}\sqrt{\frac{\log(1/x)}{1-x}}x^{a_i^\nu}(1-x)^{b_i^\nu}\exp\bigg[-\delta^\nu\frac{\bfp^2}{2\kappa^2}\frac{\log(1/x)}{(1-x)^2}\bigg],
\label{LFWF_phi}
\end{equation}
where $\kappa$ is a scale parameter comming from the soft-wall AdS/QCD model. Whit this information, it is possible to write the Dirac and Pauli form factors for spin-$\frac12$ composite particle systems \cite{Brodsky:1980zm}. In \cite{Chakrabarti:2013gra} it was found, by fitting the proton form factors from the soft/AdS/QCD model with experimental data \cite{Gay1,Gay2,Arr,Pun,Puck} that the best agreement if given with $\kappa = 0.4066$ GeV. Furthermore, in \cite{Maji:2016yqo} the flavour form factors for $u$ and $d$ in this light-front diquark model has been fitted with experimental data \cite{Cates11,Diehl13}, obtaining the value of the parameters $a^\nu_i$ and $b^\nu_i$  at the initial scale $\mu_0$, showed in Table \ref{tab_para_mu0}. \\

\begin{table}[ht]
\centering % used for centering table 
\begin{tabular}{c c c c c c}
\toprule
 \hline
$ \nu $~&~$a_1^\nu$~ & ~$b_1^\nu$~ & ~$a_2^\nu$~ & ~$b_2^\nu$~&~~$\delta^\nu$~~\\ \midrule
\hline
 ~~ $u$ ~~&~~ $0.280\pm 0.001 ~$~&~~ $0.1716\pm 0.0051$ ~~&~~ $0.84 \pm 0.02$ ~~&~~ $0.2284 \pm 0.0035$&1.0 \\ 
  $d$ & $0.5850 \pm 0.0003$ & $0.7000 \pm 0.0002$ &  $0.9434^{+0.0017}_{-0.0013}$ & $0.64^{+0.0082}_{-0.0022}$ & 1.0 \\ \bottomrule
  \hline
 \end{tabular} 
\caption{The fitted parameters  for nucleon valence $u$ and $d$ quarks at the initial scale $\mu_0$. Data from \cite{Maji:2016yqo}.} % title of Table 
\label{tab_para_mu0} % is used to refer this table in the text 
\end{table}

In the same way, using the Sachs form factors in \cite{Maji:2016yqo} is obtained the coefficients for the quark-diquark nucleon state \eqref{PS_state}: $C_S^2= 1.3872$, $C_V^2 = 0.6128$ and $C_{VV}^2 = 1$. Besides, the normalized constants $N_i^2$ are found to be:
$N_S = 2.0191, ~
N_0^{(u)} = 3.2050, ~ N_0^{(d)} = 5.9423,~
N_1^{(u)} = 0.9895, ~ N_1^{(d)} = 1.1616.$

\subsection{Quark-diquark unpolarized PDF evolution}
The unpolarized parton distribution function is defined as \cite{Bacchetta:2008af,Maji:2016yqo}
\begin{equation}
    f^{(\nu)}(x,\mu_0)=\frac{1}{2}\int \frac{d z^-}{2(2\pi)} e^{ip^+z^-/2} \langle P;S|\bar{\psi}^{(\nu)}(0)\gamma^+\psi^{(\nu)}(z^-)|P; S\rangle \bigg|_{z^+=z_T=0},
\end{equation}
which depends only on the light-cone momentum fraction $x=p^+/P^+$. Where the proton state $|P; S\rangle$ with spin $S$, is the given in Eq. (\ref{PS_state}). Indeed, $\gamma^+$ is the light-cone representation of the usual $\gamma^\mu$ matrix, detailed definition is found in \cite{Lepage:1980fj}.
The leading order QCD evolution of the unpolarized PDF is given as the standard DGLAP expansion \cite{Altarelli:1977zs, Broniowski:2007si, Maji:2016yqo}
\begin{equation}
    \int^1_0 dx x^n f(x,\mu)=\bigg(\frac{\alpha_s(\mu)}{\alpha_s(\mu_0)} \bigg)^{\gamma^{(0)}_n/2\beta_0} \int^1_0 dx x^n f(x,\mu_0), \label{DGLAP_Eq}
\end{equation}
where the anomalous dimension is determined by 
\begin{equation}
    \gamma^{(0)}_n=-2C_F\bigg(3+\frac{2}{(n+1)(n+2)}-4\sum_{k=1}^{n+1}\frac{1}{k} \bigg).
\end{equation}
And the running coupling constant is given as
\begin{equation}
    \alpha_s(\mu)=\frac{4\pi}{\beta_0 \ln(\mu^2/\Lambda^2_{QCD})}
\end{equation}
In this work we take take $C_F=4/3$, $\beta_0=9$ and  $\Lambda_{QCD}=0.226~ GeV$. The initial scale in most of the works on which ours is based is taken to be $\mu_0=0.313$ GeV, since is a value available for pion phenomenology.\\

Thus, the light-front diquark unpolarised PDFs at scale $\mu$ are given by \cite{Maji:2016yqo}
\begin{align}\label{sdiq}
    f^{(S)}(x,\mu) =  N^{2}_S(\mu)\bigg[\frac{1}{\delta^u(\mu)} x^{2a_1^u(\mu)}(1-x)^{2b_1^u(\mu)+1}+ x^{2a_2^u(\mu)-2}(1-x)^{2b_2^u(\mu)+3}\frac{\kappa^2}{(\delta^u(\mu))^2 M^2\ln(1/x)}\bigg],
\end{align}
\begin{align}\label{adiq}
    f^{(A)}(x,\mu)=&  \bigg(\frac{1}{3}N^{(\nu)2}_0(\mu)+\frac{2}{3}N^{(\nu)2}_1(\mu)\bigg)\nonumber \\
\times& \bigg[ \frac{1}{\delta^\nu(\mu)}x^{2a_1^\nu(\mu)}(1-x)^{2b_1^\nu(\mu)+1}+ x^{2a_2^\nu(\mu)-2}(1-x)^{2b_2^\nu(\mu)+3}\frac{\kappa^2}{(\delta^\nu(\mu))^2 M^2\ln(1/x)}\bigg].
\end{align}
The parameters $a^{\nu}_i,~b^{\nu}_i$ and $\delta^\nu$  are now dependent on the scale $\mu$ such that the relation \eqref{DGLAP_Eq} holds, i.e. \cite{Maji:2016yqo},
\begin{align}
    a_i^\nu(\mu)&=a_i^\nu(\mu_0) + A^\nu_{i}(\mu), \label{a_im}\\
b_i^\nu(\mu)&=b_i^\nu(\mu_0) - B^\nu_{i}(\mu)\frac{4C_F}{\beta_0}\ln\bigg(\frac{\alpha_s(\mu^2)}{\alpha_s(\mu_0^2)}\bigg),\label{b_im}\\
\delta^\nu(\mu)&= \exp\bigg[\delta^\nu_1\bigg(\ln(\mu^2/\mu_0^2)\bigg)^{\delta^\nu_2}\bigg],\label{DL}
\end{align}
where the quantities $A^\nu_{i}(\mu)$ and $B^\nu_{i}(\mu)$  are defined as 
\begin{align}
\Pi^\nu_{i}(\mu)&=\alpha^\nu_{\Pi,i} ~\mu^{2\beta^\nu_{\Pi,i}}\bigg[\ln\bigg(\frac{\mu^2}{\mu_0^2}\bigg)\bigg]^{\gamma^\nu_{\Pi,i}}\bigg|_{i=1,2} ,\label{Pi_evolu}
\end{align}
for $\Pi=A,B$. The $a_i^\nu(\mu_0)$ and $b_i^\nu(\mu_0)$ are the parameters given in Table \ref{tab_para_mu0}. It should be noted that the parameter $\delta^\mu$ tends to unity while $\mu\rightarrow \mu_0$.

In order to find the evolution parameters $\alpha_{\Pi,i}^\nu,~\beta_{\Pi,i}^\nu$, $\gamma_{\Pi,i}^\nu$ and $\delta^\nu$ it is useful to write the flavour decomposed PDFs $f^{u}(x,\mu)$ and $f^{d}(x,\mu)$. It is well discussed in \cite{Bacchetta:2008af} that for the relation between quark flavors and diquark states must have a linear behaviour with free coefficients to be determinate with experimental data. Indeed, in the same way the proton state \eqref{PS_state} has to be consistent with the real world under the same coefficients $C_S,~C_V$ and $C_{VV}$, which was how the flavored form factors were decomposed from the diquarks and such parameters found in \cite{Maji:2016yqo}. So, the flavour decomposed PDFs are given as
\begin{align}
    f^{u}(x,\mu)&=C_S^2 f^{(S)}_1(x,\mu)+C_V^2 f^{(V)}_1(x,\mu),\\
f^{d}(x,\mu)&=C_{VV}^2 f^{(VV)}_1(x,\mu).
\end{align}
Then, the flavoured PDF $f^{\nu}(x,\mu$) in the light-front quark-diquark model can be written as
\begin{align}
    f^\nu(x,\mu)=&  N^{(\nu)}\bigg[\frac{1}{\delta^\nu(\mu)} x^{2a_1^\nu(\mu)}(1-x)^{2b_1^\nu(\mu)+1}+ x^{2a_2^\nu(\mu)-2}(1-x)^{2b_2^\nu(\mu)+3}\frac{\kappa^2}{(\delta^\nu(\mu))^2 M^2\ln(1/x)}\bigg],\label{Eq_xf1mu}
\end{align}
where $N^{(u)}=(C_S^2N_s^2+C^2_V(\frac{1}{3} N_0^{(u)2}+\frac{2}{3} N_1^{(u)2}))$ and $N^{(d)}=C^2_{VV}(\frac{1}{3} N_0^{(d)2}+\frac{2}{3} N_1^{(d)2})$ for $u$ and $d$ quarks respectively.\\

In this work, we have followed the fashion of \cite{Maji:2016yqo} and we have obtained the values of the evolution parameters by fitting the flavour PDFs \ref{Eq_xf1mu} with data from NNPDF2.3 QCD+QED NNLO  \cite{Ball:2013hta}. The fit was performed in gnuplot \cite{gnuplot}, an open source plotting tool using non-linear least-square theory, taking first a $f^\nu$ depending on parameters $\Pi^\nu_{i}(\mu)$, then getting the evolution parameters $\alpha_{\Pi,i}^\nu,~\beta_{\Pi,i}^\nu$, $\gamma_{\Pi,i}^\nu$ and $\delta^\nu$.
The unpolarized  PDF  data  was  fitted for 100 equal-spaced data points for $x \in (0,1)$ and $\mu^2=2,4,8,16,32,64, 128, 256~ GeV^2$. The fitted parameters are shown in Table \ref{tab_evopar} for $\alpha_{\Pi,i}^\nu,~\beta_{\Pi,i}^\nu$ and $\gamma_{\Pi,i}^\nu$, while in Table \ref{tab_DL}  are shown the fitted $\delta^\nu$. In appendix \ref{appa} we show the different fits performed for the scales mentioned above.\\

\begin{figure}[htbp]
\begin{minipage}[c]{0.98\textwidth}
\small{(a)}\includegraphics[width=8.0cm,clip]{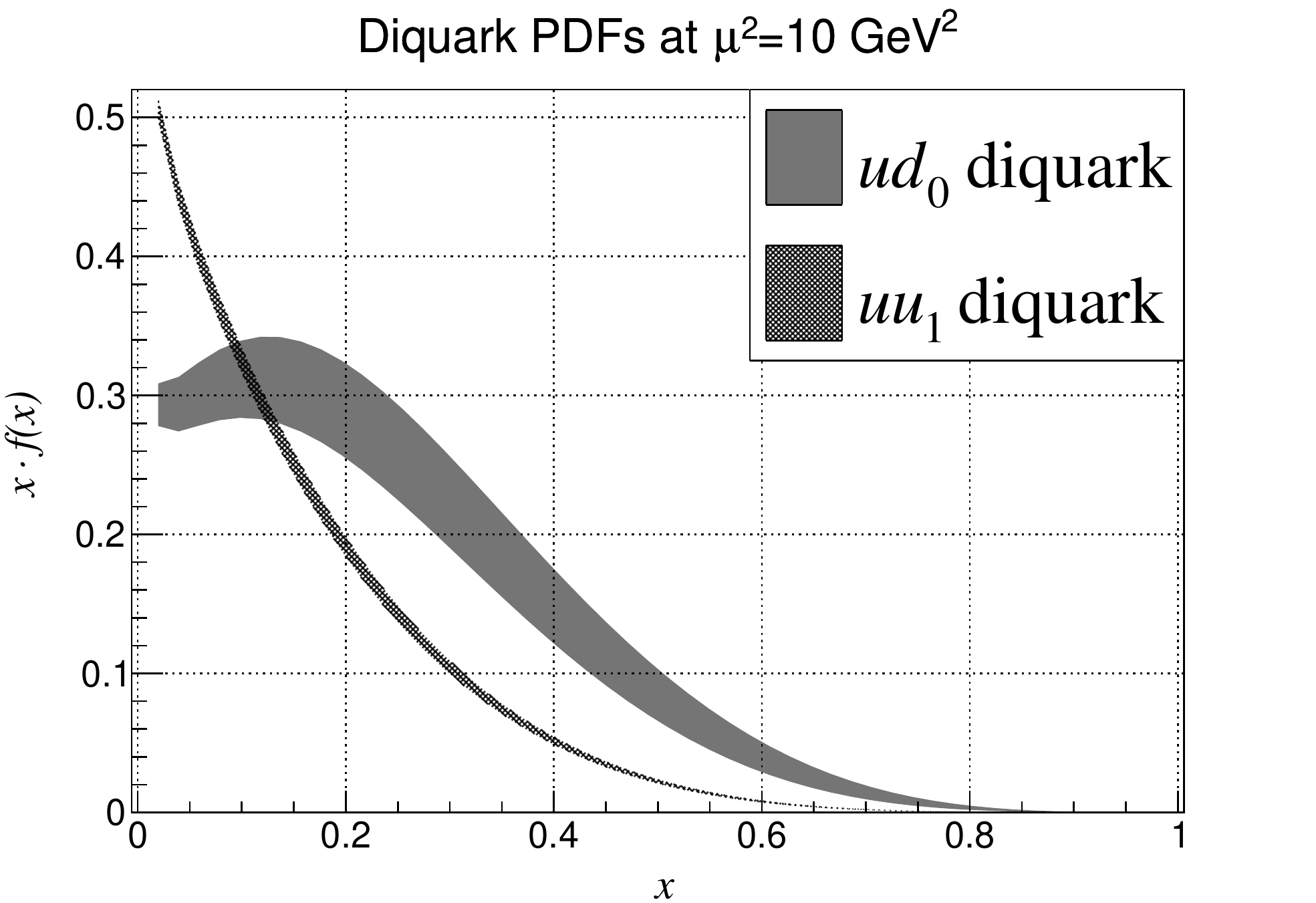}
\small{(b)}\includegraphics[width=8.0cm,clip]{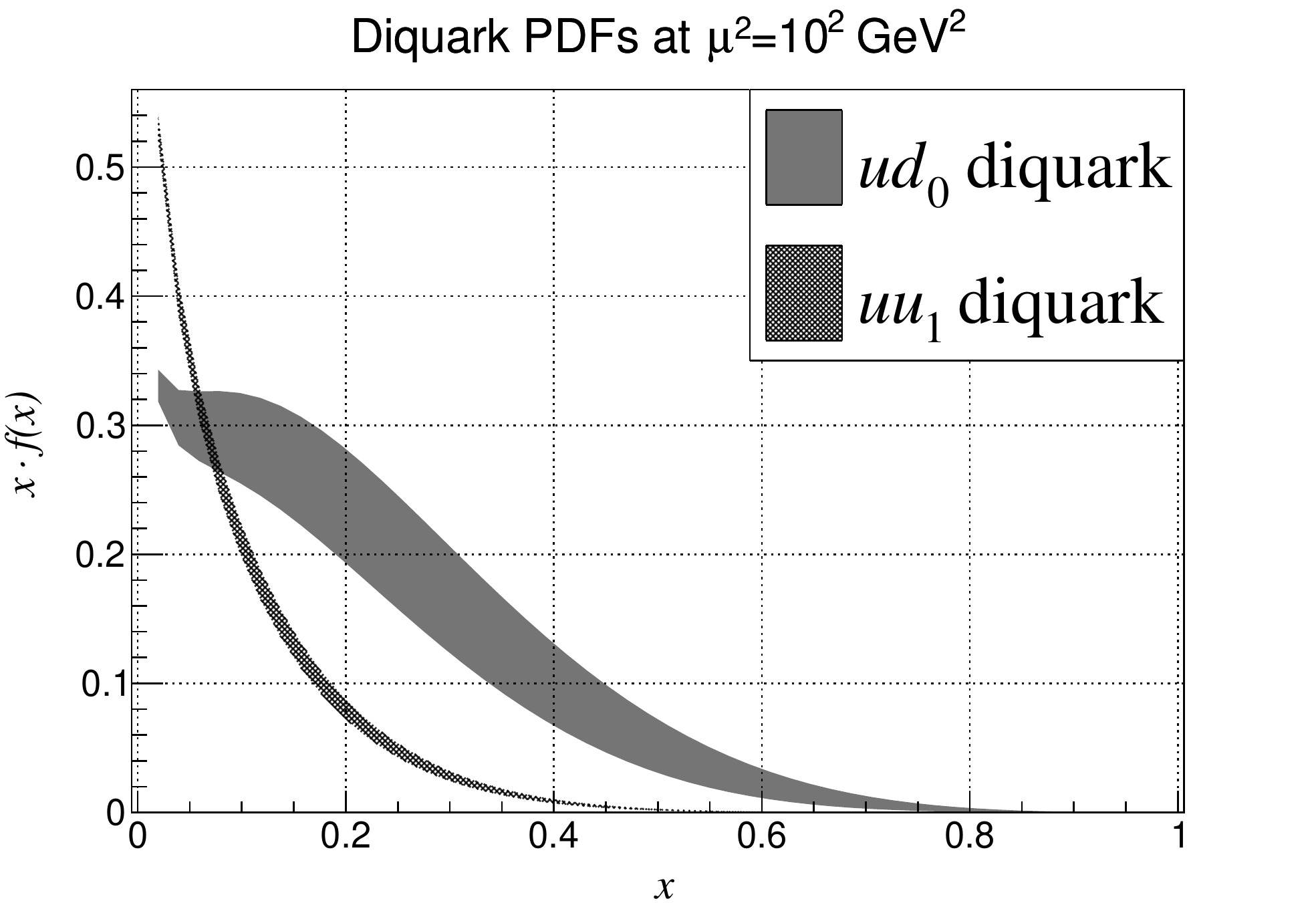}
\end{minipage}
\begin{minipage}[c]{0.98\textwidth}
\small{(c)}\includegraphics[width=8.0cm,clip]{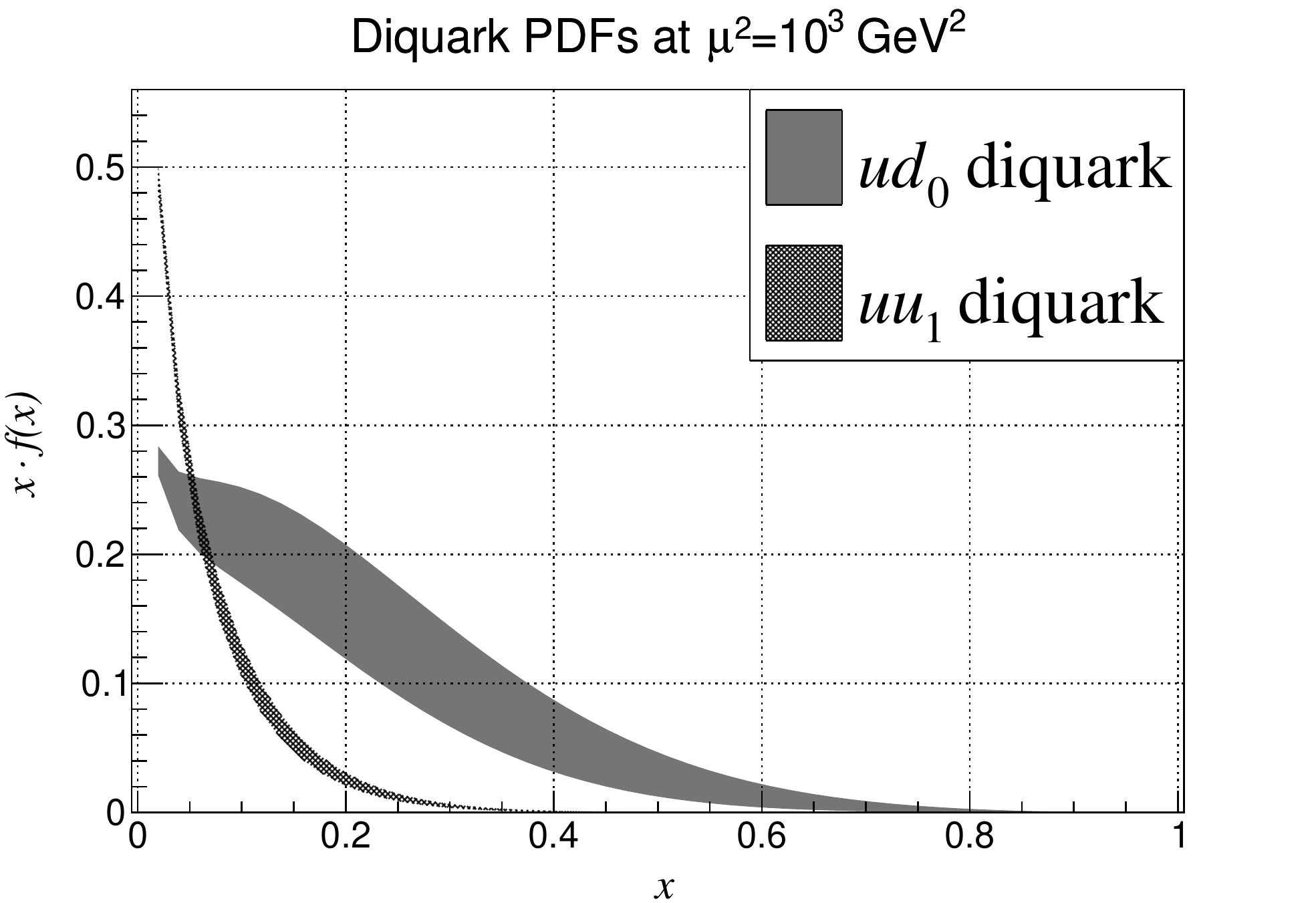}
\small{(d)}\includegraphics[width=8.0cm,clip]{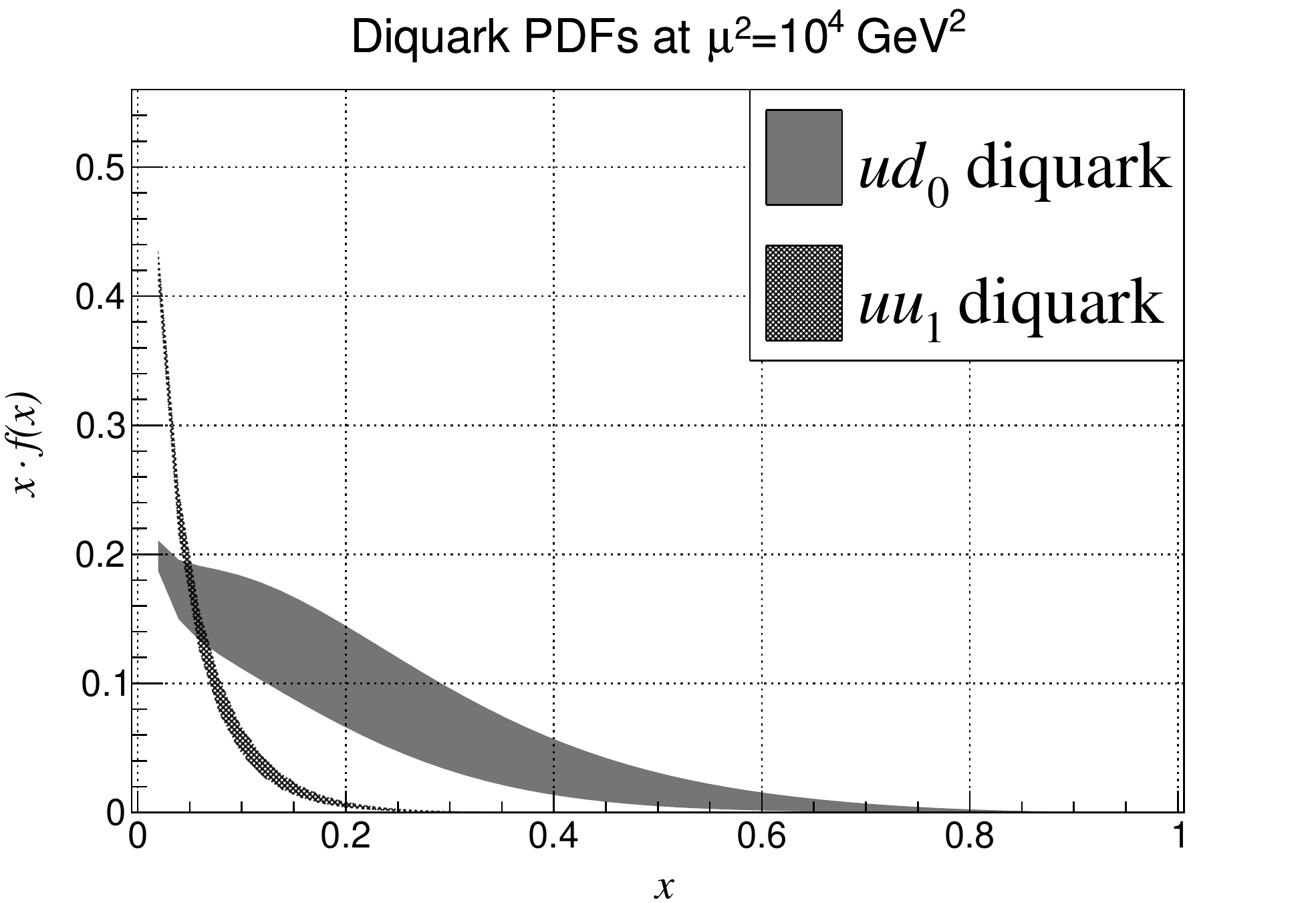}
\end{minipage}
\caption{\label{fig:pdfdi} Figures for $x\cdot f^{(S)}(x)$ and $x\cdot f^{(VV)}(x)$ at different scale energies. In (a), (b), (c) and (d) it is shown the behaviour of diquark PDFs for energies $\mu^2=10, 10^2, 10^3$ and $10^4$ GeV respectively. In smooth bands  the case of the scalar diquark, while the checkered bands are for the isovector-vector diquark.} % the data of Table.\ref{tab_ABd} are fitted by varying evolution parameters $\alpha^\nu_{P,i},\beta^\nu_{P,i} ~and~ \gamma^\nu_{P,i}$, for $d$ quark.}
\end{figure}

\begin{table}[ht]
\centering % used for centering table 
\begin{tabular}{c c c c c}
\toprule
 \hline
 $\Pi_i^\nu(\mu)$~~&~~$\alpha_i^\nu$~~&~~$\beta_i^\nu$~~ & ~~$\gamma_i^\nu$~~ & ~~$\chi^2/d.o.f$~~  \\ 
 \midrule
 \hline
$A_1^u$ &~~ $-0.196314\pm0.002266$ ~~&~~ $-0.197209\pm0.01021$ ~~&~~ $0.927163\pm0.03627$ ~~&~~ 0.09\\
$B_1^u$ & $6.4894\pm0.04592$ & $0.161127\pm0.006494$ & $-0.910813\pm0.02185$ &0.17\\ 
$A_2^u$ & $-0.441651\pm0.002674$ & $-0.0389503\pm0.005802$ & $0.306214\pm0.01902$ & 0.995\\ 
$B_2^u$ & $2.58149\pm0.26410$ & $-0.0548368\pm0.07806$ & $-0.807298\pm0.2779$ &1.54\\ 
$A_1^d$ &~~ $-0.119059\pm0.002517$ ~~&~~ $-0.124819\pm0.018800$ ~~&~~ $0.952914\pm0.0601$ ~~&~~ 0.27\\
$B_1^d$ & $12.8481\pm0.09134$ & $0.0976609\pm0.006134$ & $-0.80035\pm0.01510$ &0.53\\
$A_2^d$ & $-0.514816\pm0.000724$ & $-0.001555\pm0.001244$ & $0.171831\pm0.003307$ & 0.41\\ 
$B_2^d$ & $1.10727\pm0.00703$ & $0.084447\pm0.005591$ & $-0.5719\pm0.01486$ &0.03\\ \bottomrule
 \hline
 \end{tabular} 
\caption{PDF evolution parameters with 95\% confidence bounds. Using data from NNPDF2.3 QCD+QED NNLO  \cite{Ball:2013hta}} % title of Table 
\label{tab_evopar} % is used to refer this table in the text 
\end{table}
\begin{table}[ht]
\centering  
\begin{tabular}{c c c c }
\toprule
 \hline
 $\delta^\nu(\mu)$~~~&~~~$\delta_1^\nu$~~~~&~~~~~$\delta_2^\nu$~~~ & ~~$\chi^2/d.o.f$~~  \\ \midrule
 \hline
$\delta^u$ & $0.35074\pm0.03009$ & $0.48314\pm0.06732$ & 10.5\\
$\delta^d$ & $0.406762\pm0.007024$ & $~0.46990\pm0.01275$ & 3.79 \\ \bottomrule
\hline
\end{tabular} 
\caption{PDF evolution parameter $\delta^\nu_1$ and $\delta^\nu_2$ for $\nu=u,d$. Using data from NNPDF2.3 QCD+QED NNLO  \cite{Ball:2013hta} }  
\label{tab_DL}  
\end{table}
With this data applied to the PDFs \eqref{sdiq} and \eqref{adiq}, we have drawn the functions ($x\cdot f(x)$) of the isoscalar-scalar diquark and isovector-vector diquark for energies $\mu^2=10, 10^2, 10^3$ and $10^4$ GeV$^2$ shown in Fig. \ref{fig:pdfdi} (a), (b), (c) and (d) respectively. The smooth bands show the case of the scalar diquark, while the checkered bands are for the mentioned vector diquark. It is important to note that $\frac{1}{3} N_0^{(u)2}+\frac{2}{3} N_1^{(u)2} \approx N^2_S$ from values reported in \cite{Maji:2016yqo}, so the behaviour of $f^{(S)}$ (isoscalar-scalar) and $f^{(V)}$ (isoscalar-vector) is very close.
%we have the issue of PDF for diquarks covered. Now we will deal with cross sections for diquarks.

\section*{Conclusions}
The soft-wall light front AdS/QCD has allowed us to construct Parton Distritution Functions for diquarks in agreement with the data observed for quarks in experiments. We have particularly taken data from NNPDF2.3 QCD+QED NNLO  \cite{Ball:2013hta}, but the model can be adapted to desired experimental data with $u$ and $d$ quark PDF information. Although the uncertainties for the values in $\Pi_i^\nu(\mu)$ reported here should be still improved, an acceptable fit for the functions \eqref{sdiq} and \eqref{adiq} is shown in our parameters in Tabs. \ref{tab_evopar} and \ref{tab_DL} looking at $\chi^2/d.o.f.$

In general terms, the behaviour of diquark PDFs observed in Figs. \ref{fig:pdfdi} reveals a similarity to the quark PDFs. Such behaviour goes in the sense that as the energy scale increases, a shift to $ x = 0 $ of the peak of the functions is visible; as well as, while $ x $ approaches 1, $ xf $ tends to vanish exponentially. This fact can be compared in \cite{Maji:2016yqo}, where using the same model with  NNPDF21(NNLO) \cite{DelDebbio:2007ee} data have been fitted the $u$ and $d$ quark PDFs. 

The phenomenological diquark PDFs here reported are intended to be tested within the framework of particle collisions. Especially for us, it is expected to study the effect on the production of hadrons in collision simulations of the AdS/QCD quark-diquark nucleon model taking into account participant diquarks in hard processes.
\subsection*{Acknowledgements}
B.R. would like to thank to the Secrearía Nacional de Ciencia y Tecnología (Ref. Grant FINDECYT/EDUCA CTi 02-2019) of Guatemala for financial support. Also many thanks to Anatoly Egorov for many valuable comments and  insightful discussions.
\bibliographystyle{unsrt}  
%\bibliography{references}  %%% Remove comment to use the external .bib file (using bibtex).
%%% and comment out the ``thebibliography'' section.

%%% Comment out this section when you \bibliography{references} is enabled.

\appendix
\section{Parameter fitting for PDF evolution from NNPDF2.3 QCD+QED NNLO\label{appa}}
\begin{figure}[htbp]
\begin{minipage}[c]{0.98\textwidth}
\small{(a)}\includegraphics[height=5.0cm,clip]{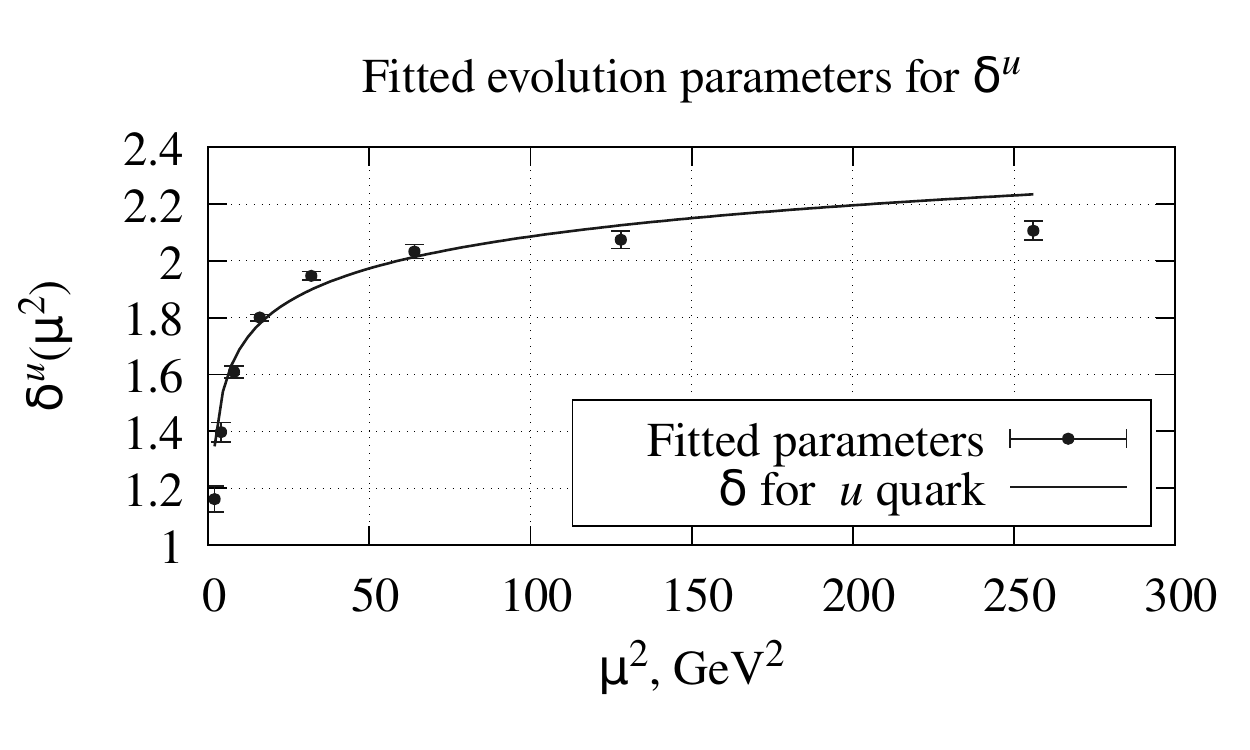}
\small{(b)}\includegraphics[height=5.0cm,clip]{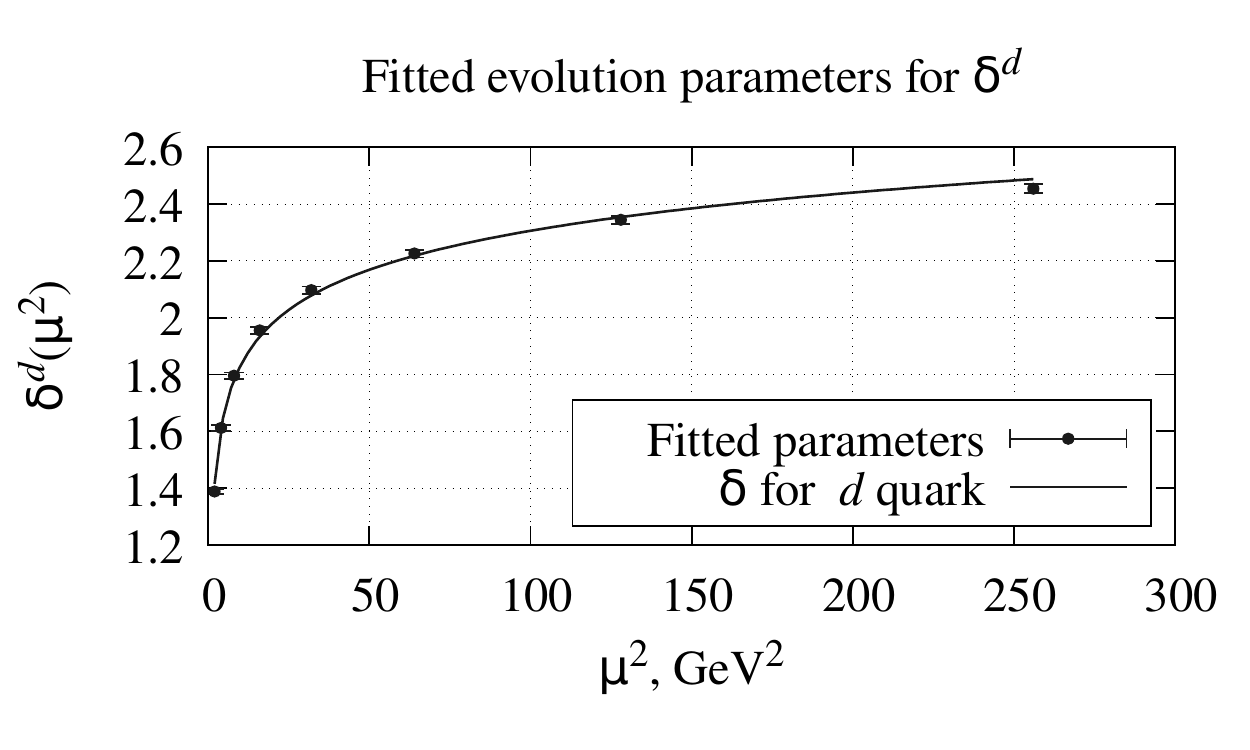}
\end{minipage}
\caption{\label{fig_del}  The data of Table.\ref{tab_DL} are fitted by varying evolution parameters $\delta^\nu_1$ and $\delta^\nu_2$ in dependence of the energy scale, for (a)  $u$  quark and (b)  $d$ quark.}
\end{figure}

\begin{figure}[htbp]
\begin{minipage}[c]{0.98\textwidth}
\small{(a)}\includegraphics[width=7.5cm,clip]{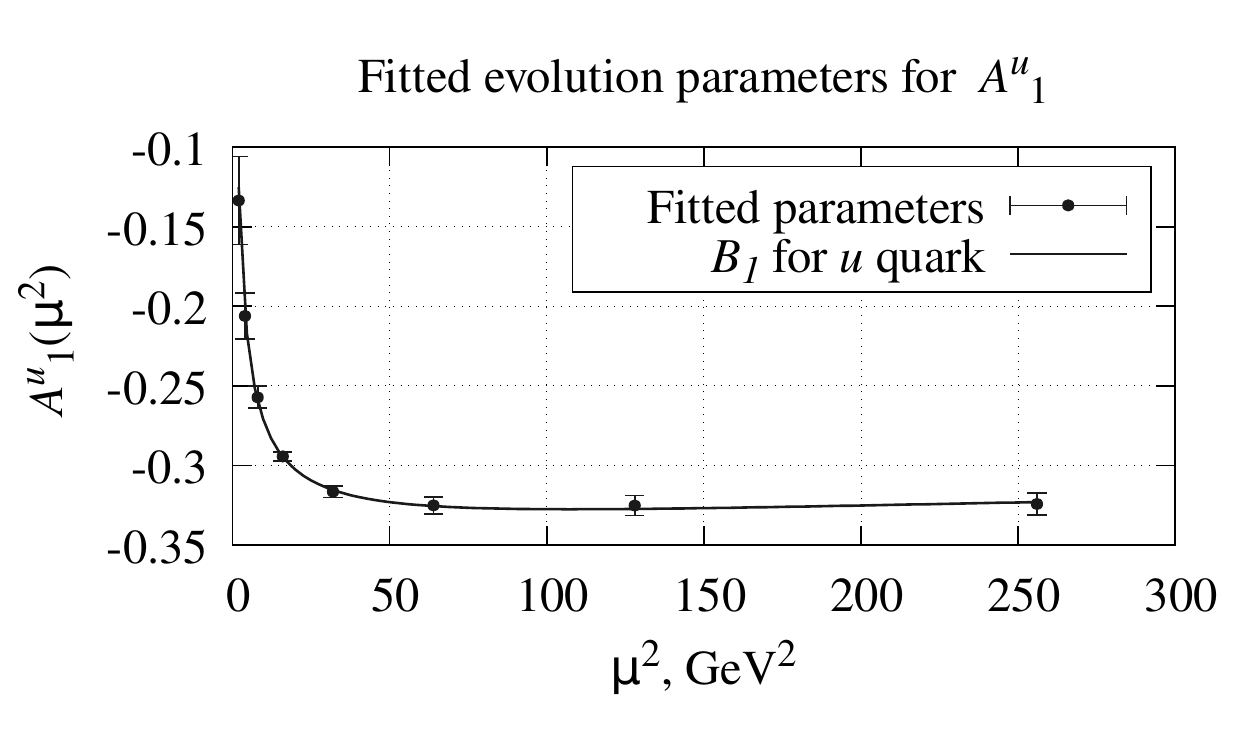}
\small{(b)}\includegraphics[width=7.5cm,clip]{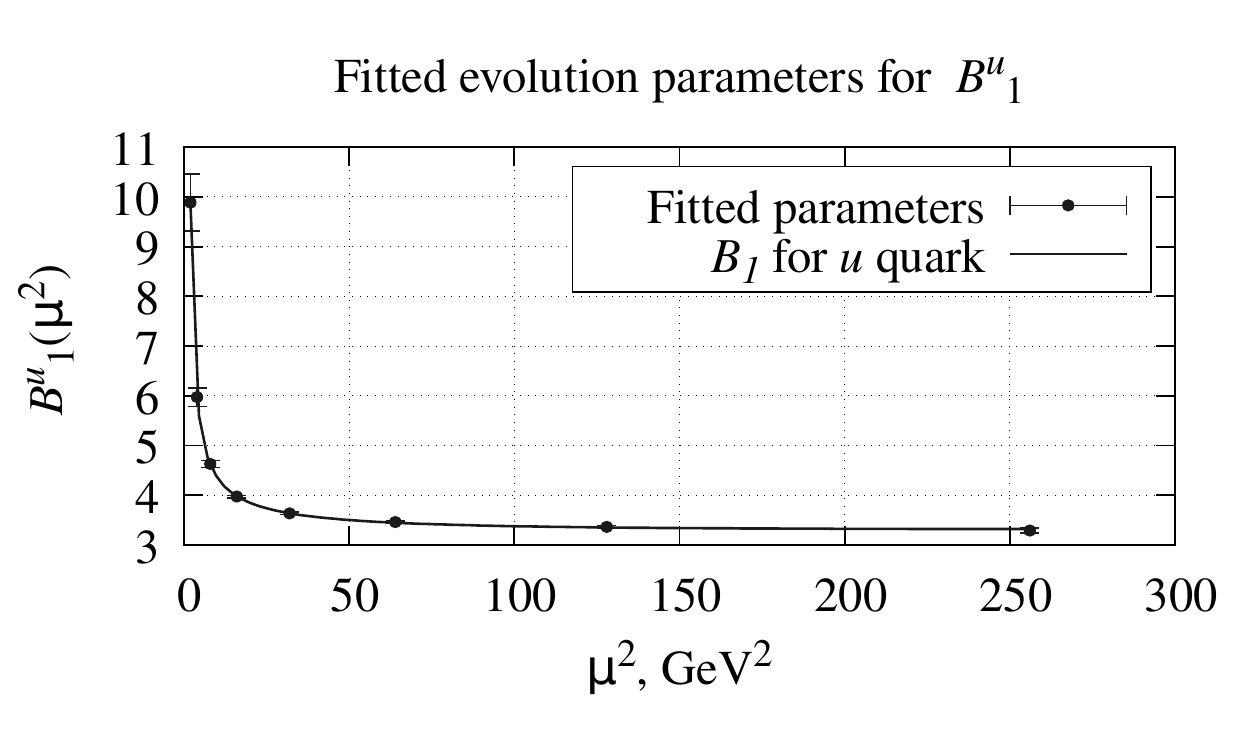}
\end{minipage}
\begin{minipage}[c]{0.98\textwidth}
\small{(c)}\includegraphics[width=7.5cm,clip]{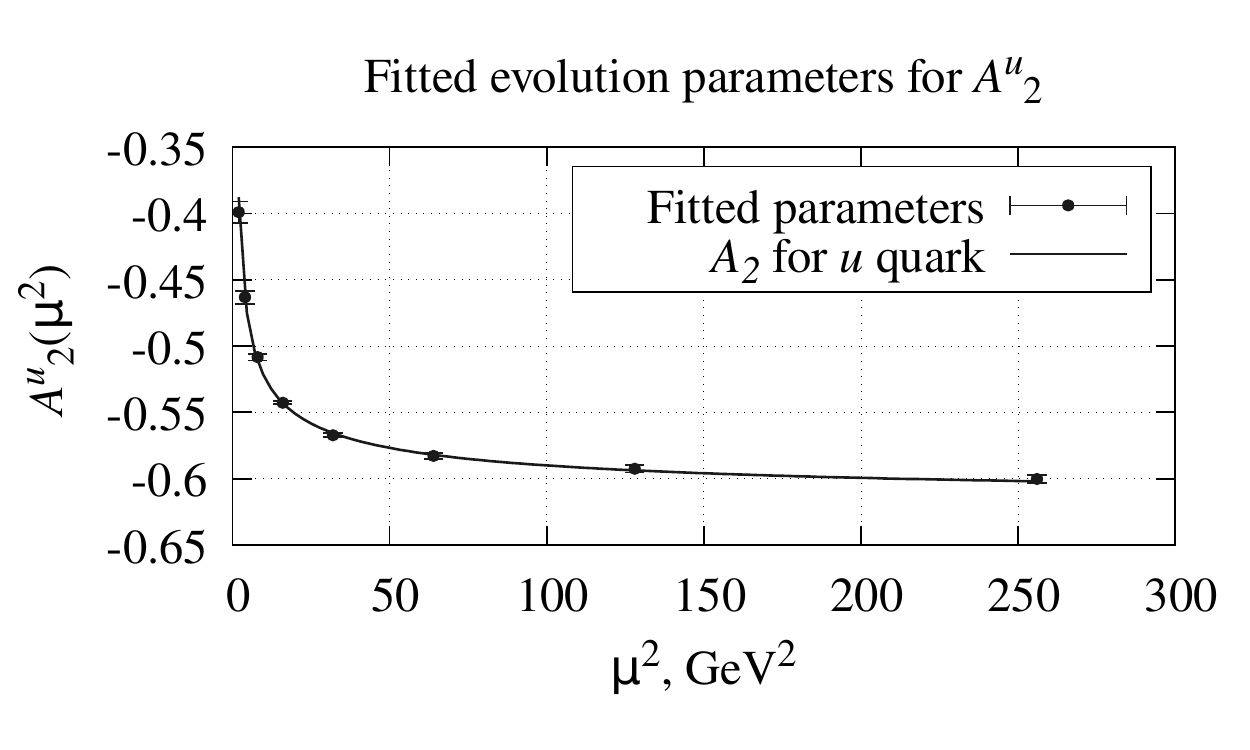}
\small{(d)}\includegraphics[width=7.5cm,clip]{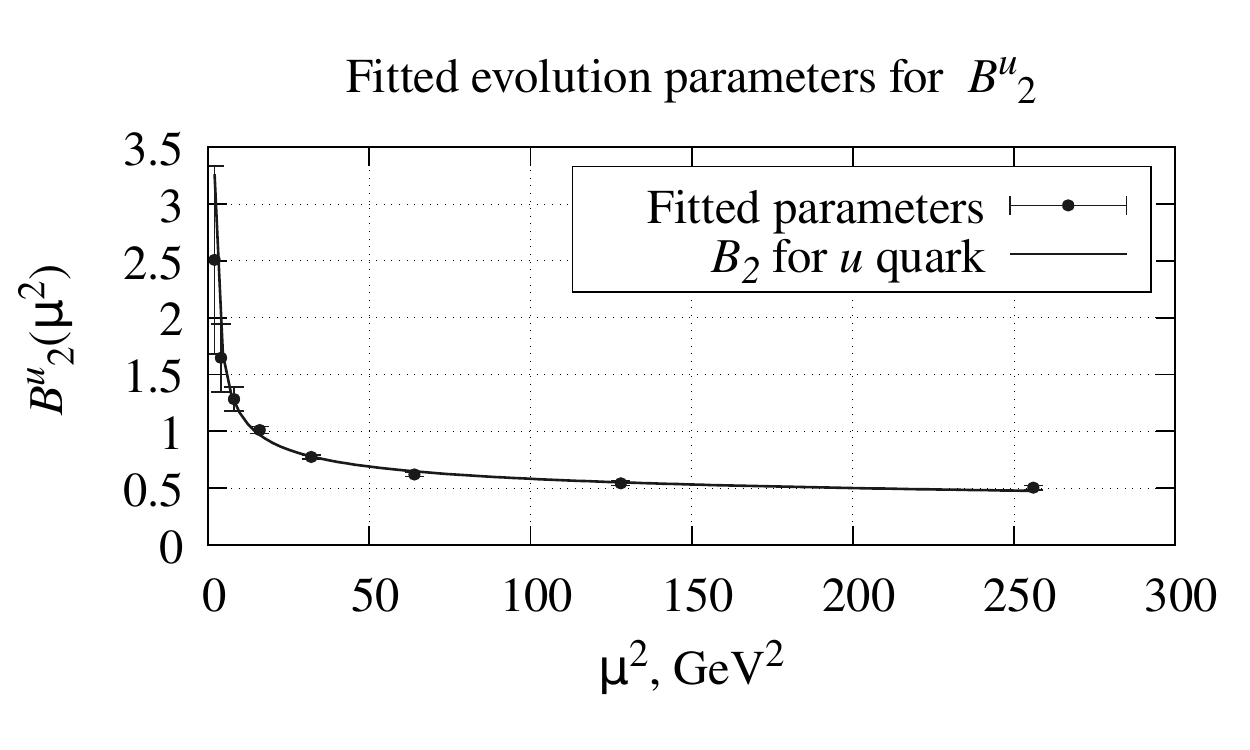}
\end{minipage}
%\caption{\label{fig_ABu} Using Eq.(\ref{Pi_evolu}), the data of Table.\ref{tab_ABu} is fitted by varying evolution parameters $\alpha^\nu_{P,i},\beta^\nu_{P,i} ~and~ \gamma^\nu_{P,i}$, for $u$ quark.}
%\end{figure}
%\begin{figure}[htbp]
\begin{minipage}[c]{0.98\textwidth}
\small{(e)}\includegraphics[width=7.5cm,clip]{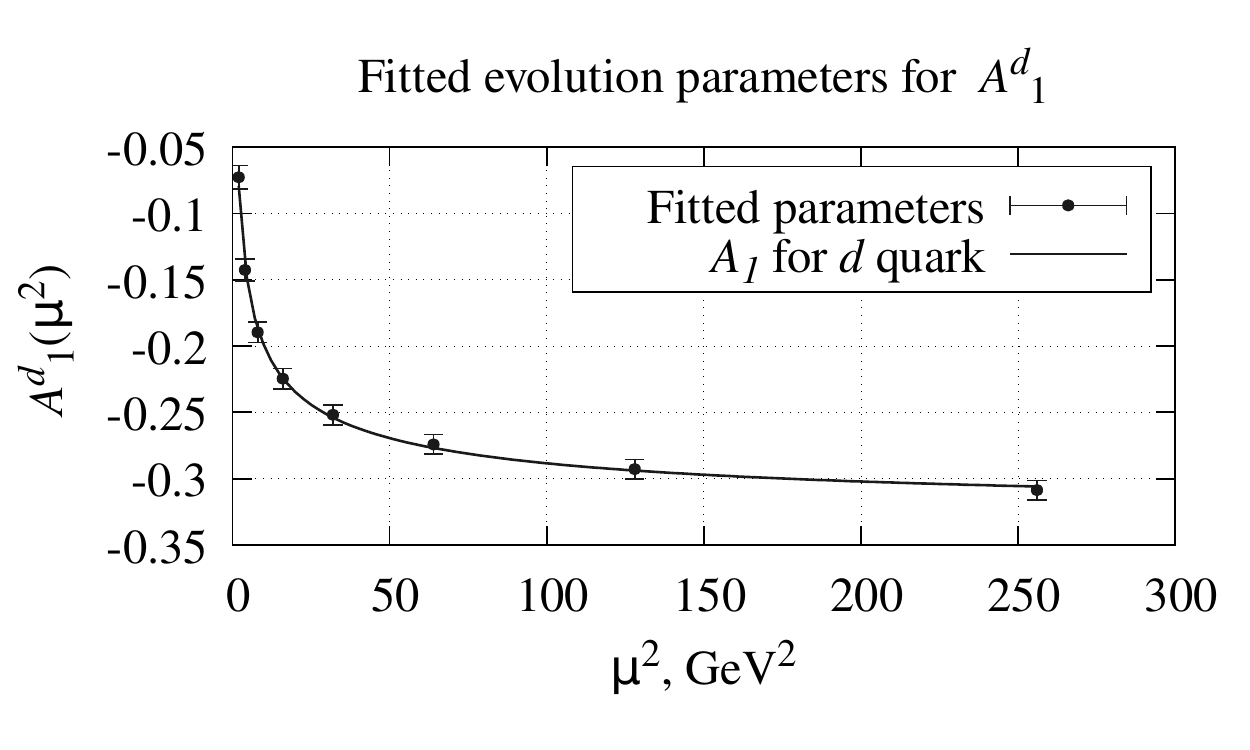}
\small{(f)}\includegraphics[width=7.5cm,clip]{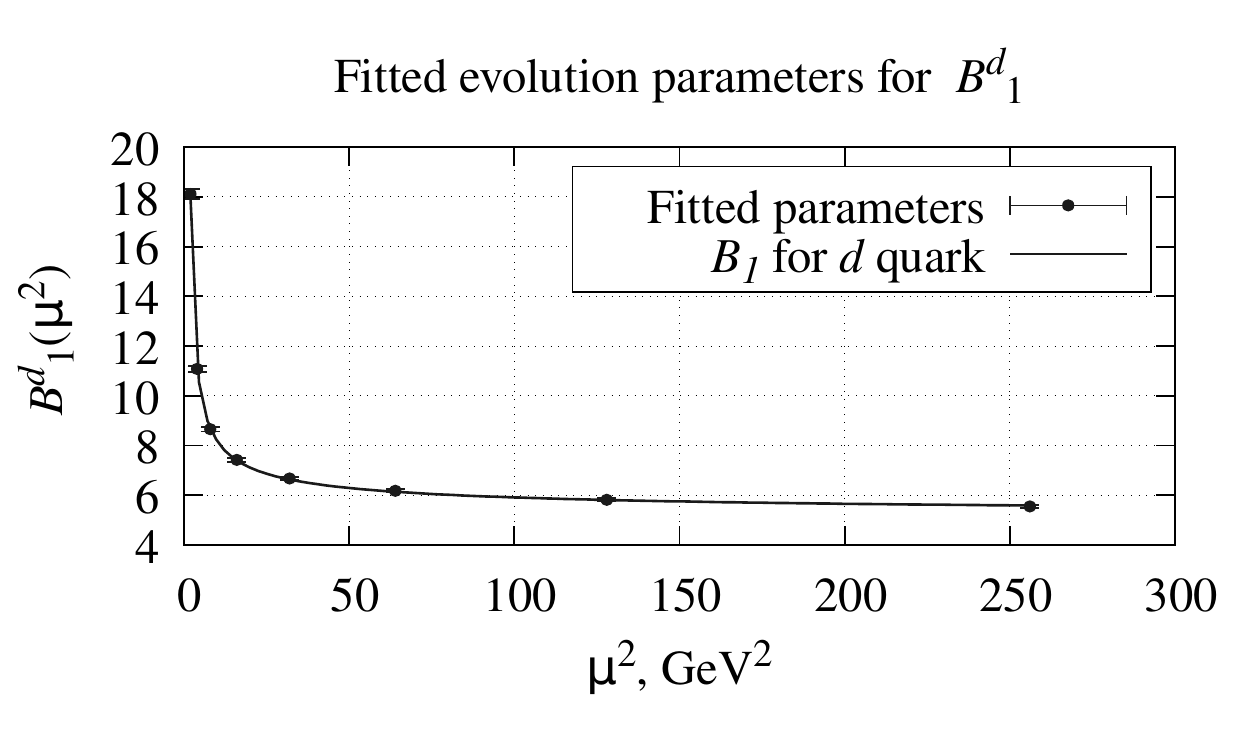}
\end{minipage}
\begin{minipage}[c]{0.98\textwidth}
\small{(g)}\includegraphics[width=7.5cm,clip]{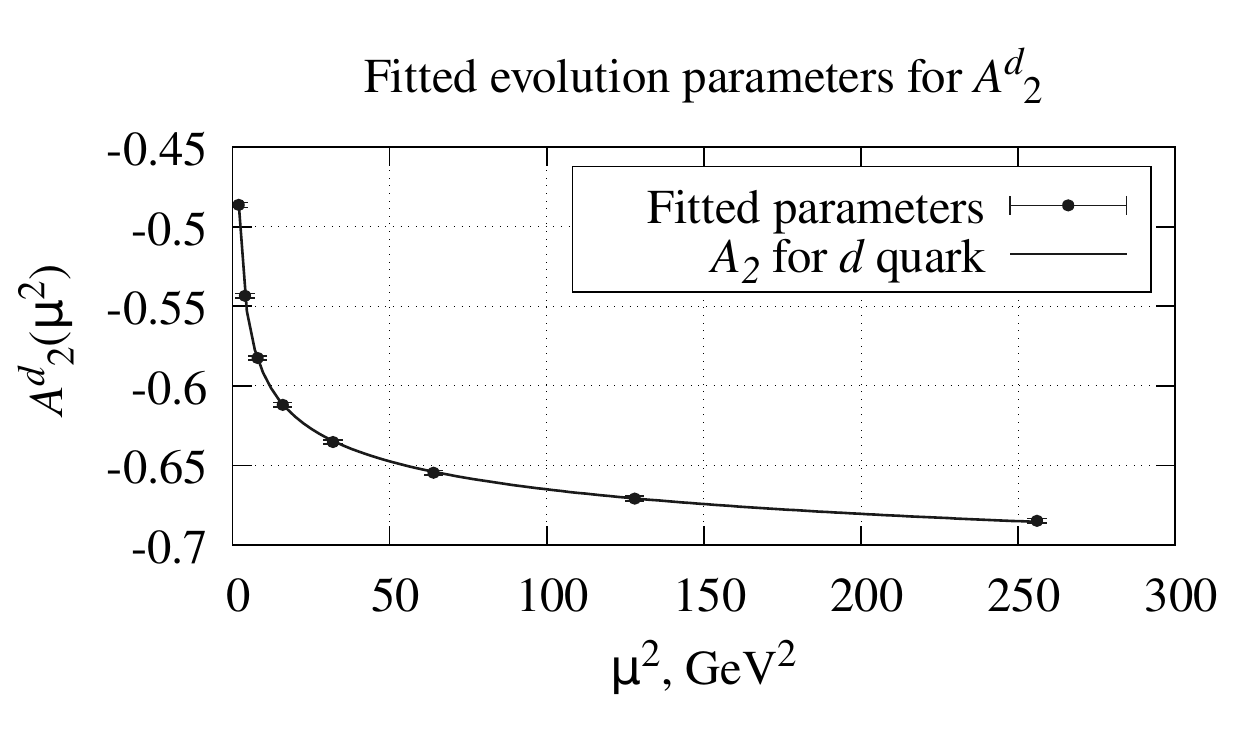}
\small{(h)}\includegraphics[width=7.5cm,clip]{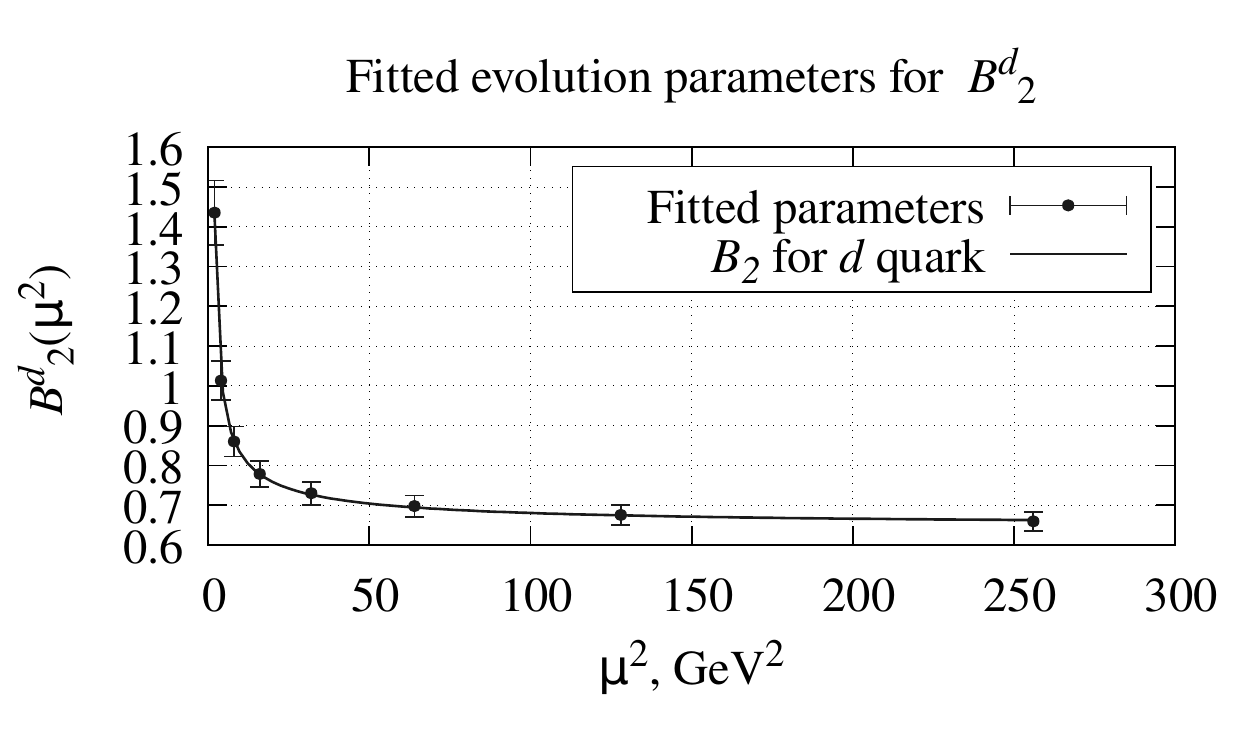}
\end{minipage}
\caption{\label{fig_ABud} Using Eq.(\ref{Pi_evolu}), the data of Table.\ref{tab_ABu} are fitted by varying evolution parameters $\alpha^\nu_{\Pi,i},\beta^\nu_{\Pi,i}$ ~and~ $\gamma^\nu_{\Pi,i}$ in dependence of the energy scale, for $u$ quark ($(a) - (d)$). Similar data fitting plots for $d$ quark(Table.\ref{tab_ABd}) are shown 
in $(e) - (h)$.} % the data of Table.\ref{tab_ABd} are fitted by varying evolution parameters $\alpha^\nu_{P,i},\beta^\nu_{P,i} ~and~ \gamma^\nu_{P,i}$, for $d$ quark.}
\end{figure}

%%%%%%%%%%%%%%%%%%%%%%%%%%%
The scale evolution of  $A_i^\nu$ and $B_i^\nu$ is parameterized by $\alpha_i^\nu,~\beta_i^\nu$, and $\gamma_i^\nu$. While $\delta^\nu$ is parameterized by $\delta_1^\nu$ and $\delta_2^\nu$.  $f(x,\mu)$ is given by Eq.(\ref{Eq_xf1mu}) along with Eqs.(\ref{a_im},\ref{b_im},\ref{DL}) and Eq.(\ref{Pi_evolu}).
$f(x,\mu)$ dependent of the parameters $A_i^\nu,~ B_i^\nu$ and $\delta^\nu$ are fitted at 8 different scales $\mu^2$ in Table \ref{tab_ABu} for $u$ quark, while the fitted parameters for $d$ quark are given Table \ref{tab_ABd}.  
%The last column in each  table indicates the least $\chi^2$  error in the PDF estimation. 
Each  $\chi^2/d.o.f$ has been evaluated for 100 equally-spaced points for different $x\in (0,1)$
%i.e., for the five parameter fit, we have  (100-5)=95 degrees of freedom. 
The fitting of the parameters at $\mu^2=2,4,8,16,32,64,128$ and $256$ GeV$^2$ are shown in Fig.\ref{fig_ABud} and Fig.\ref{fig_del}. The data points are extracted from NNPDF2.3 QCD+QED NNLO  \cite{Ball:2013hta}. It should be noted that the $\chi^2/d.o.f$ values show that the uncertainty ranges found with the fit are over-estimated, this is because for this first instance, uncertainties were not taken from the PDF data of \cite{Ball:2013hta}. An improvement in this fact is expected for future works.
% using Eq.(\ref{Eq_xf1mu}) and Eqs.(\ref{a_im},\ref{b_im},\ref{DL}). 
%The error bars shown in the plots are the errors in  the extracted values of the  parameters  due to the uncertainties in the PDF data.

\begin{table}[ht]
\centering % used for centering table 
 \scriptsize{
\begin{tabular}{c c c c c c c }
\toprule
 \hline
% \footnotesize{
 $\mu^2~GeV^2$~~&~~$A_1^u$~~ & ~~$B_1^u$~~ & ~~$A_2^u$~~ & ~~$B_2^u$~~ & ~~$\delta^u$~~ &~~$\chi^2/d.o.f$~~ \\ \midrule
 \hline
2.0   & -0.133482 $\pm$ 0.02763  & 9.88657 $\pm$ 0.57650 & -0.398994 $\pm$ 0.008142 & 2.50897  $\pm$ 0.82540 & 1.16148 $\pm$ 0.04635 & 9.15238e-06 \\
4.0   & -0.206116 $\pm$ 0.01457  & 5.97471 $\pm$ 0.18860 & -0.463197 $\pm$ 0.00477  & 1.64702  $\pm$ 0.29730 & 1.39743 $\pm$ 0.03360 & 2.73019e-06 \\
8.0   & -0.257193 $\pm$ 0.006954 & 4.63066 $\pm$ 0.06874 & -0.508357 $\pm$ 0.002519 & 1.28388  $\pm$ 0.10670 & 1.60899 $\pm$ 0.02080 & 6.44055e-07 \\
16.0  & -0.294376 $\pm$ 0.002982 & 3.97296 $\pm$ 0.02359 & -0.542694 $\pm$ 0.001198 & 1.01137  $\pm$ 0.03058 & 1.80059 $\pm$ 0.01118 & 1.22395e-07 \\
32.0  & -0.316551 $\pm$ 0.003503 & 3.63544 $\pm$ 0.02254 & -0.567223 $\pm$ 0.001527 & 0.774017 $\pm$ 0.02007 & 1.94722 $\pm$ 0.01540 & 1.77367e-07 \\
64.0  & -0.325091 $\pm$ 0.005228 & 3.45959 $\pm$ 0.02855 & -0.582866 $\pm$ 0.002362 & 0.620872 $\pm$ 0.01846 & 2.03299 $\pm$ 0.02475 & 4.2913e-07  \\
128.0 & -0.325202 $\pm$ 0.006335 & 3.36176 $\pm$ 0.03040 & -0.592571 $\pm$ 0.002883 & 0.542005 $\pm$ 0.01836 & 2.07403 $\pm$ 0.03057 & 7.16693e-06 \\
256.0 & -0.32426  $\pm$ 0.006969 & 3.28934 $\pm$ 0.04935 & -0.600304 $\pm$ 0.003162 & 0.504157 $\pm$ 0.01947 & 2.10605 $\pm$ 0.03384 & 6.94007e-06\\ \bottomrule
\hline
 \end{tabular} 
\caption{Fitting of the PDF $f(x)$ at various scales for $u$ quark.  } % title of Table 
\label{tab_ABu} % is used to refer this table in the text 
}
\end{table} 
\begin{table}[ht]
\centering % used for centering table 
 \scriptsize{
\begin{tabular}{c c c c c c c } 
\toprule
 \hline
 $\mu^2~GeV^2$~~&~~$A_1^d$~~ & ~~$B_1^d$~~ & ~~$A_2^d$~~ & ~~$B_2^d$~~ & ~~$\delta^d$~~&~~$\chi^2/d.o.f$ \\ \midrule
 \hline
2.0   & -0.0726864 $\pm$ 0.008883 & 18.1153 $\pm$ 0.19740 & -0.486242 $\pm$ 0.001573 & 1.43562  $\pm$ 0.08108 & 1.38805 $\pm$ 0.008694 & 1.54472e-06 \\
4.0   & -0.142581  $\pm$ 0.008094 & 11.0790 $\pm$ 0.11820 & -0.54338  $\pm$ 0.001465 & 1.01348  $\pm$ 0.04888 & 1.61203 $\pm$ 0.009794 & 1.42185e-06 \\
8.0   & -0.189572  $\pm$ 0.007766 & 8.65931 $\pm$ 0.09223 & -0.582432 $\pm$ 0.001427 & 0.860135 $\pm$ 0.03799 & 1.79567 $\pm$ 0.01090  & 1.35531e-06 \\
16.0  & -0.224539  $\pm$ 0.007578 & 7.42345 $\pm$ 0.07912 & -0.611889 $\pm$ 0.00141  & 0.778533 $\pm$ 0.03241 & 1.95505 $\pm$ 0.01194  & 1.29922e-06 \\
32.0  & -0.25185   $\pm$ 0.007459 & 6.67682 $\pm$ 0.07126 & -0.635289 $\pm$ 0.001402 & 0.729967 $\pm$ 0.02896 & 2.09703 $\pm$ 0.01291  & 1.24493e-06 \\
64.0  & -0.274124  $\pm$ 0.007376 & 6.17642 $\pm$ 0.06598 & -0.654558 $\pm$ 0.001398 & 0.697843 $\pm$ 0.02660 & 2.22594 $\pm$ 0.01381  & 1.19305e-06 \\
128.0 & -0.292732  $\pm$ 0.007316 & 5.81723 $\pm$ 0.06219 & -0.670823 $\pm$ 0.001397 & 0.675414 $\pm$ 0.02486 & 2.34434 $\pm$ 0.01466  & 1.14381e-06 \\
256.0 & -0.30861   $\pm$ 0.007272 & 5.54758 $\pm$ 0.05934 & -0.684823 $\pm$ 0.001397 & 0.659191 $\pm$ 0.02353 & 2.45415 $\pm$ 0.01547  & 1.09728e-06 \\ \bottomrule
\hline
 \end{tabular}
\caption{ Fitting of the PDF $f(x)$ at various scales for $d$ quark. \label{tab_ABd}.} % title of Table 
 % is used to refer this table in the text 
 }
\end{table}

\end{document}